\documentclass{article}
\usepackage[left=2.5cm,right=2cm,top=1.5cm,bottom=2.5cm]{geometry}
\usepackage{graphics}

\begin{document}
\title{Highly stable piezoelectrically tunable optical cavities}

\author{Katharina M\"ohle, Evgeny V. Kovalchuk, Klaus D\"oringshoff, Moritz Nagel, Achim Peters\\
\\
\normalsize Institut f\"ur Physik, Humboldt-Universit\"at zu Berlin, Newtonstr. 15, 12489 Berlin, Germany, \\
\normalsize  fax: +49 30 2093 4718, email: katharina.moehle@physik.hu-berlin.de}      
\date{\small Version 07.02.2013\\
\small accepted for publication in Applied Physics B. The final publication is available at link.springer.com, doi:10.1007/s00340-012-5322-0}

\maketitle
\begin{abstract}
We have implemented highly stable and tunable frequency references using optical high finesse cavities which incorporate a piezo actuator. As piezo material we used ceramic PZT, crystalline quartz, or PZN-PT single crystals. Lasers locked to these cavities show a relative frequency stability  better than $1\cdot10^{-14}$, which is most likely not limited by the piezo actuators. The piezo cavities can be electrically tuned over more than one free spectral range ($>\, 1.5\,$GHz) with only a minor decrease in frequency stability. Furthermore, we present a novel cavity design, where the piezo actuator is prestressed between the cavity spacer components. This design features a hermetically sealable intra cavity volume suitable for, e.g., cavity enhanced spectroscopy. 
\end{abstract}
\section{Introduction}

Frequency stabilized lasers are widely used in, e.g., atomic physics, precision metrology, or molecular spectroscopy. The stabilization is usually implemented using either a TEM$_{00}$ Eigenmode of an optical cavity \cite{DHK83} or an atomic or molecular transition \cite{HMT99} as a reference. In both methods, only discrete laser frequencies, determined by the specific reference, can be selected. For an optical cavity, the resonance frequencies are separated by the free spectral range (FSR) given by FSR $= c/(2L)$, where $c$ is the speed of light and $L$ the cavity length. Thus (when using a TEM$_{00}$ cavity mode), the laser frequency can only be stabilized to frequencies separated by multiples of the FSR, which is typically on the order of $1\,$GHz.

For stabilization to arbitrary frequencies within the laser tuning range, different techniques can be employed, including the application of acousto-optical modulators (AOM), serrodyne modulation \cite{KVB12}, or an additional offset frequency locked laser \cite{YeH99}. Further options are the use of an unequal arm-length interferometer \cite{KJL09,JKL10,SHD10} or a sideband locking scheme \cite{TnL08}. These methods mostly require components with high power consumption or complex additional hardware. We, thus, have investigated a simpler approach, namely to tune the resonance frequency of the cavity by changing the cavity length itself with a piezo actuator. 

We analyzed high finesse piezo-tunable cavities in the context of the proposed space based gravitational wave observatory LISA \cite{LISA11,DaR03}, where a tunable frequency prestabilization is required within the proposed three-level noise reduction scheme. Furthermore, highly stable piezo-tunable cavities could find applications in cavity enhanced spectroscopy \cite{MYD99,FSM08}, as highly stable transfer cavities \cite{BJM79,PAM06,RAS10}, or as optical local oscillators in atomic and molecular spectroscopy \cite{LIE04,JLL11}.

The best cavity stabilized lasers have currently a relative frequency stability in the low $10^{-16}$ range \cite{JLL11,KHG12,YCI99}, limited by thermal noise \cite{NKC04}. They are typically made of glass materials which have a low coefficient of thermal expansion (CTE) and feature special vibration insensitive designs \cite{MMM09,WOP08,LHN07} to suppress all length fluctuations. 

Our aim was to investigate the achievable stability of a piezo-tunable cavity, where the influence of the mechanical and thermal properties of the actuator and the voltage noise in the tuning voltage have to be considered. In addition, one has to deal with piezo effects like hysteresis and creep.

Piezo-tunable cavities are already commonly used for laser resonators or as transfer cavities \cite{LSD91}. A piezo-tunable cavity where the cavity spacer is compressed with an extrinsic actuator \cite{CRM03} has been implemented with stabilities in the $10^{-13}$ range, allowing, however, only a few $10\,$kHz tuning range. Furthermore, it has been demonstrated that a laser system including stabilization to a cavity with an intrinsic piezo actuator can be realized with a few kHz linewidth \cite{LIE04}. However, the real potential and limitations of tunable high finesse cavities have so far not been thoroughly investigated.

In this paper, we now present results on the frequency stability and tuning capabilities of four piezo cavities built with different piezo actuators. After describing the experimental setup, we report on the frequency stability which is achieved with short circuited piezo actuators, revealing the intrinsic stability of the piezo cavities without any voltage related effects. Then, we discuss the behavior of the cavities with applied voltage as well as the performance of a piezo cavity locked to another reference. In the final part of the paper, we present a novel piezo cavity design with integrated piezo actuators.

\section{Experimental setup}

\subsection{Cavities}

\begin{figure}
		\centering\resizebox{0.3\textwidth}{!}{%
		\includegraphics{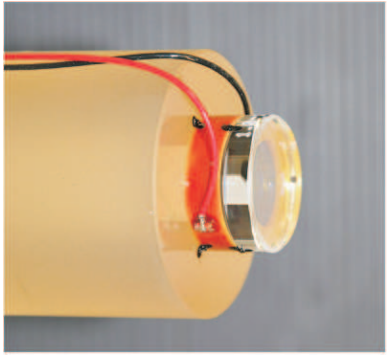}}
		\caption{Picture of the cavity with PZT stack actuator attached between spacer and mirror via a three point glue joint on each side (black spots).}
		\label{fig:piezo}
	\end{figure}

The cavities are made of a $10\,$cm long Zerodur spacer, one or two piezo actuators, and one plane and one concave ($50\,$cm radius of curvature) fused silica mirror. The mirrors are coated for high reflectivity at $1064\,$nm with a targeted finesse of 300\,000. The piezo actuators are attached between mirror and spacer by three gluing points on each side (Fig. \ref{fig:piezo}) to facilitate disassembly of the cavities after characterization. It was thus possible to test different piezo materials in a cavity configuration using the same spacer. We also used a `fixed' cavity with both mirrors optically contacted to an identical spacer, which can be used to characterize the setup.

Four different piezo actuators were tested: a stack actuator and a tube actuator made from lead zirconate titanate (PZT), a quartz crystal, and a lead zinc niobate-lead titanate (PZN-PT) single crystal. Two different \linebreak types of PZT actuators were selected as we expected differences in the mechanical stability of a stack actuator, which consists of many PZT and electrode layers, and a tube actuator, where the transversal piezo effect is used.

The properties of the piezo materials which are relevant for the use in a cavity are given in Table \ref{tab:piezo materials}. The piezoelectric coefficient $d_{ij}$ is a measure for the potential displacement of the actuator. All materials have a more than two orders of magnitude higher CTE than cavity spacer materials like Zerodur or ULE, which may influence the long-term performance of a cavity with an incorporated piezo actuator. The Young modulus ($E$), which should be high for a low vibration sensitivity as well as for a low thermal noise level, is of the same order of magnitude than that of Zerodur or ULE. The mechanical quality factor ($Q$), which should be high to achieve a low thermal noise level, shows significant differences between the piezo materials as well as in comparison to Zerodur (3\,100) and ULE (61\,000) \cite{NKC04}.

    \begin{table}
    \centering
{\small    
    \caption{\small Relevant properties of the used piezo materials PZT (PI ceramics, PIC 151), quartz (Boston Piezo-Optics, crystal quartz) and PZN-PT (Microfine Materials Technologies, PZN-(6-7)\%PT single crystals). $d_{ij}$ denotes the  piezoelectric coefficient, CTE the coefficient of thermal expansion, E the Young Modulus and Q the mechanical quality factor. For the PZN-PT single crystal no CTE value could be found in the literature; the given value is derived from measurement. It should be noted that the listed values describe only the pure piezo material and do not account for adhesive or electrodes as used in the case of the PZT stack actuator. For these reasons the stack actuator actually has a CTE of $+3.5 \cdot 10^{-6}$/K.}
    \label{tab:piezo materials}
        	\begin{tabular}{l *{4}c}
    	\hline\noalign{\smallskip}
      Material  & $d_{ij}$  & CTE	 & E 			& Q\\
      					& (pm/V)				   & ($10^{-6}$/K) & (GPa) 	&\\
      \noalign{\smallskip}\hline\noalign{\smallskip}
      PZT 	 & $d_{33}$: 500  & -5    & 60   & 100 		 \\
       			 & $d_{31}$: -210 & 		  & 		 & 		 		 \\
      quartz & $d_{11}$: -2.3 & 13	  & 76.5 & $200\,000$ \\
      PZN-PT & $d_{33}$: 2000 & $<10$ & 62.5 & 100      \\ 
      \noalign{\smallskip}\hline       
      \end{tabular}
      }     
 		\end{table}

Each piezo actuator allows a different tuning range, depending on the piezoelectric coefficient of the material and in the case of the PZT stack and PZT tube actuator on the dimensions of the actuator. Except for the quartz piezo cavity, all piezo cavities can be tuned over more than one free spectral range, which implies the highest degree of flexibility, since the laser can be stabilized to every frequency in the laser tuning range. The measured tuning coefficients and tuning ranges are shown in Table \ref{tab:piezo cavities}.
 
	\begin{table}
	\centering
	\small{
   \caption{\small Specifications and tuning properties of the different piezo cavities and the optically contacted fixed cavity analyzed in this work. All values derived from measurement are marked with an asterisk. $L_c$ denotes the total cavity length (including actuators and for some cavities Zerodur adapter plates) and $L_p$ the length of the used piezo actuator, whereas the factor 2 indicates the use of two actuators. FSR is the free spectral range and $\delta \nu$ the linewidth of the optical cavity resonance. The thermal noise limit $S_{tn}$ of the whole cavity is calculated using the formulas in \cite{NKC04}. $\sigma_{tn}/\nu_0$ describes the thermal noise floor in the relative Allan deviation. For the calculation of the effective CTE of the assembled cavities the contribution of the adhesive was neglected. The maximum displacement values $\Delta\nu_{max}$ are calculated under the condition that voltages from $-200\,$V to $+1000\,$V can be applied to the actuators. The displacement values for the quartz piezo are only theoretical values, since the piezo effect was covered by another effect, presumably electrostatic, and no tuning coefficients could be measured.}
    \label{tab:piezo cavities}
    \centering
    \begin{tabular}{l *{11}c}
	    \hline\noalign{\smallskip}
cavity	&$L_c$  &$L_p$	&FSR	&$\delta\nu$ &coupling	&$S_{tn}$		&$\sigma_{tn}/\nu_0$ &cavity	&$\Delta\nu/U$	&$\Delta\nu_{max}$\\

				&  			&				&			& 					&	efficiency&$@\,1\,$Hz	&	 &CTE 	 	&								&\\
		
				&(mm) 	&(mm)	  &(GHz)&(kHz)			 & 	(\%)				&(Hz/$\sqrt{\textrm{Hz}}$) &$\times 10^{-16}$ 	&($10^{-6}$/K) &(MHz/V)			 &(GHz)	\\
	      \noalign{\smallskip}\hline\noalign{\smallskip}
PZT stack	&107	  &7			   		&1.40	 &37 			    &20*		 	&$0.17$									 	 &7.15								& 0.35				 & 7.5*		 	&9 	\\ 
PZT tube 	&130    &$10 \cdot 2$ &1.15	 &6*  			 	&80*		 	&$0.23$									 	 &9.64								& -0.75				 & 12*			&14.4	\\
PZN-PT		&110	  &3					 	&1.30	 &40* 			 	&18*		 	&$0.19$									 	 &7.81								& 0.3*				 & 2.2*			&2.64	\\
quartz		&110	  &$5 \cdot 2$	&1.36	 &13  			 	&50*			&$0.12$									 	 &4.92								& 1.2					 & 0.012 		&0.014\\
fixed			&100	  &-					 	&1.50	 &32* 			 	&25			  &$0.13$									 	 &5.51								& 0.02				 & -				&- 		\\
			\noalign{\smallskip}\hline         
    \end{tabular}
}
 	\end{table}

\subsection{Test environment}

To minimize vibrational sensitivity, the cavities are mounted in their Airy points on Viton O-rings. They are placed inside two thermal shields that serve as thermal low pass filters with measured time constants of $9\,$h and $15\,$h. The shields are mounted on an optical breadboard inside a vacuum chamber, which is evacuated to $2\cdot10^{-6}\,$mbar with a 360 l/s turbo pump. The vacuum chamber is supported by four pneumatic vibration isolators. We use Nd:YAG NPRO lasers (Lightwave Electronics, LW122 and LW124) at $1064$\,nm and the Pound-Drever-Hall (PDH) technique \cite{DHK83} for frequency stabilization. The laser light is attenuated such that $20\,\mu$W impinges on the cavities and guided into the vacuum chamber by a polarization maintaining single mode fiber. The lenses and mirrors for mode matching and coupling into the cavities, as well as the PDH photo detectors, are located on the optical breadboard inside the vacuum chamber. 

The frequency noise is analyzed by a beat note measurement between a laser stabilized to the tunable piezo cavity and a laser stabilized to another high finesse reference cavity. The reference cavity was either a cavity made of ultra low expansion (ULE) glass, which is vertically mounted and stabilized to its zero expansion temperature \cite{DMN10} or a fused silica (FS) cavity \cite{HSM09}. Since the frequency noise of these references is almost one order of magnitude lower than the noise of the piezo cavities, their noise contribution to the beat note can be neglected. The beat note is recorded with frequency counters (Stanford Research Systems SR620 and Pendulum CNT91).

\section{Frequency stability}
  
\subsection{Short circuited piezo actuator}

\begin{figure*}
		\centering\resizebox{0.98\textwidth}{!}{%
		\includegraphics{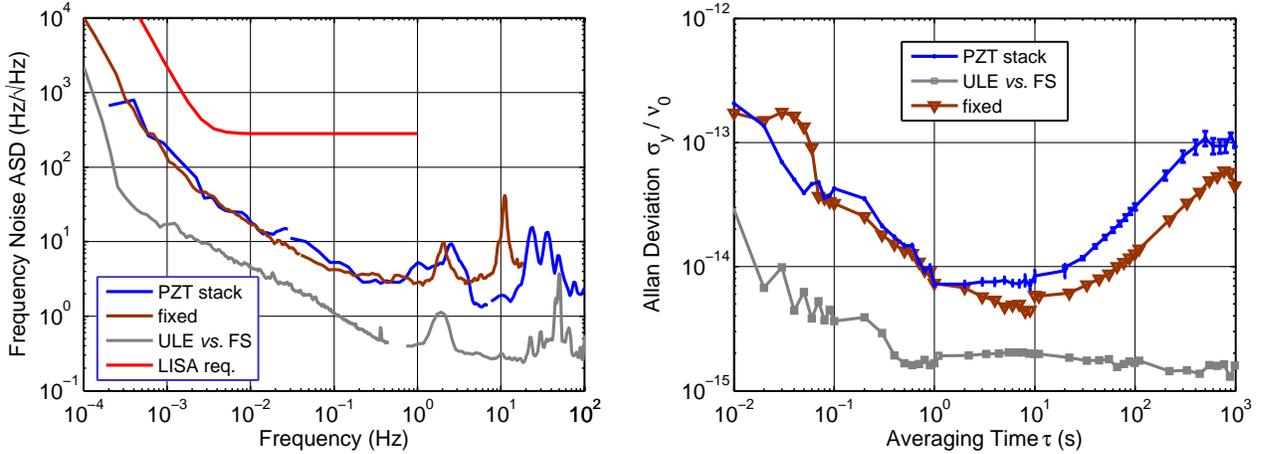}}
		\caption{ASD and relative Allan deviation of the pzt stack cavity measured versus the ULE reference. The frequency noise of the `fixed' cavity and that of the beat measured between the ULE reference cavity and the fused silica (FS) reference are shown as well. The red line in the noise spectrum describes the frequency noise requirements for a tunable prestabilization in the LISA measurement band ($0.1\,$mHz and $1$\,Hz). For the calculation of the ASD and Allan deviation a quadratic drift is removed in each case to reveal the intrinsic stability of the cavities. (Without this removal the oscillations of the air conditioning system are limiting.) Typical error bars are shown for the Allan deviation of the PZT stack cavity.}
		\label{fig:noise_pztstack}
	\end{figure*}

\begin{figure*}
		\centering\resizebox{0.98\textwidth}{!}{%
		\includegraphics{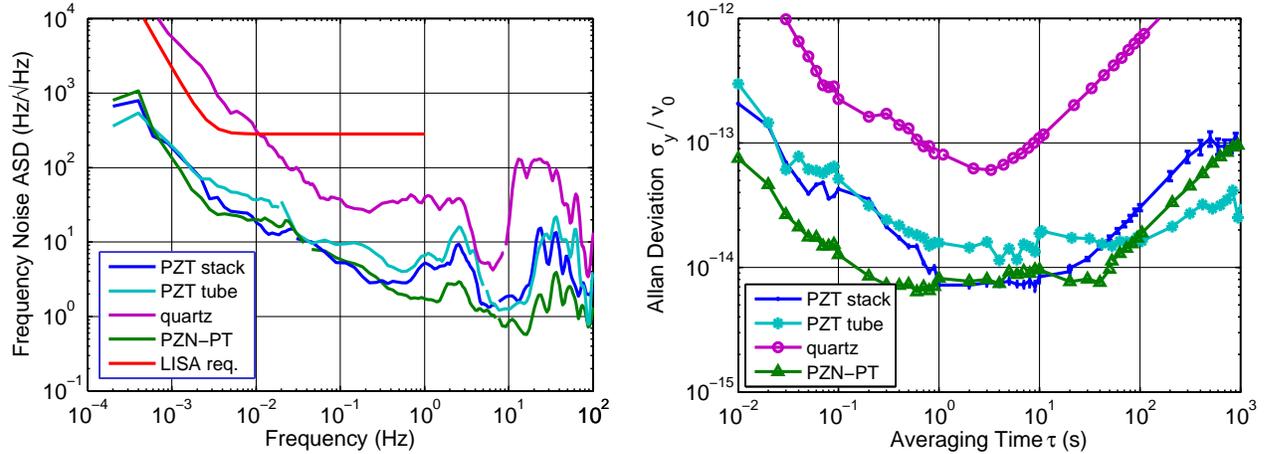}}
		\caption{ASD and relative Allan deviation of all tested piezo cavities. All measurements were made versus the ULE reference, and again a quadratic drift was removed in each case.}
		\label{fig:noise_all}
	\end{figure*}

To determine the intrinsic stability of the piezo cavities without any voltage-related effects, the frequency noise of the different piezo cavities was measured with short circuited piezo actuators. Figure \ref{fig:noise_pztstack} exemplarily shows the amplitude spectral density (ASD) and the Allan deviation of the pzt stack cavity in comparison to the fixed cavity and the ULE reference. A summary of the frequency noise curves of all tested piezo cavities is shown in Fig. \ref{fig:noise_all}.
	
All piezo cavities except for the quartz piezo cavity show similar noise characteristics. At Fourier frequencies above 1 Hz the noise spectrum is dominated by seismic noise, while between $4\,$mHz and $1\,$Hz the performance is limited by flicker noise and below $4\,$mHz by random walk noise. Typical drift rates are $100\,$Hz/s, mainly caused by residual slow temperature changes of the cavities inside the thermal shields.

The best performance is reached with the PZN-PT cavity, which shows a frequency noise below $30\,\textrm{Hz}/\sqrt{\textrm{Hz}}$ for Fourier frequencies above $3\,$mHz and a relative Allan deviation below $1 \cdot 10^{-14}$ at integration times from $0.2\,$s to $50\,$s. The cavities with PZT stack and PZT tube actuator show a slightly increased noise level, mainly due to a higher sensitivity to seismic noise. This can be attributed to small differences in the gluing technique, which was improved over time. 

The measurements with the `fixed' cavity were performed in the same setup where the piezo cavities were analyzed. A comparison of the performance of the `fixed' cavity with those of the piezo cavities shows that the flicker and random walk noise levels are approximately the same. We, thus, assume that in the corresponding frequency range the frequency stability of the piezo cavities is not limited by the piezo actuator but by the experimental setup. It is likely that the piezo cavities are, in principle, capable of achieving an even better performance.

To characterize the experimental setup, the sensitivity of the system to various external influences was determined. The sensitivity to intensity changes was identified to be $2\,$MHz/W corresponding to a fractional frequency change of $7\cdot 10^{-9}$/W. A tilt sensitivity of $7\,$Hz/$\mu$rad ($2.5\cdot10^{-14}$/$\mu$rad) was measured by slowly releasing the pressure in two adjacent pneumatic isolators and comparing the beat record with the tilt measurements. A vibration sensitivity of $3\,$MHz/g ($1.1\cdot10^{-8}$/g) could be estimated by comparing the seismic peaks in the beat measurement with seismometer measurements, and a sensitivity to pressure changes of $50\,$MHz/mbar ($1.8\cdot10^{-7}$/mbar) could be observed by simultaneously measuring the pressure and beat during a pressure rise in the vacuum chamber (all values for PZT stack cavity). The corresponding noise levels were measured and scaled with the sensitivity. None of these effects could be identified as to be limiting in the low frequency range. The calculated thermal noise limit, which for the piezo cavities is only slightly higher than for the fixed cavity is also not yet reached (Table \ref{tab:piezo cavities}).

We observed that within the measurement sensitivity of a few femtometer, the Eccobond 285 epoxy did not provide sufficient stability after the specified curing time of $24\,$h. When a piezo cavity was placed in vacuum shortly after the specified curing time, 10-12 weeks were necessary to reach the noise levels presented in this paper. Heating of the assembly under vacuum conditions did not significantly accelerate the curing process. Therefore, the cavities assembled later were cured at normal pressure and elevated temperatures ($50^\circ$C) to reduce the curing time to less than one week.

The frequency noise of the quartz piezo cavity is much higher than that of the other piezo cavities, though probably not due to its intrinsic properties. The quartz piezo cavity could not be tuned as expected since the piezo effect was covered by another unwanted effect, which we assume to be electrostatic. Therefore, the cavity was disassembled after only a few weeks of measurement. As a consequence measurements were only done when the adhesive was not yet fully cured and the frequency noise was still limited by adhesive related processes.

\subsection{Applying a constant piezo driving voltage}

It is evident that noise in the voltage driving the piezo actuator could decrease the stability of the cavity and the corresponding laser frequency. Frequency noise estimations can easily be done using the tuning coefficients listed in Table \ref{tab:piezo cavities}. A higher tuning coefficient requires a lower driving voltage to reach the same tuning range, but does also increase the sensitivity to voltage noise. In the case of the PZT stack cavity, e.g., a voltage noise below $270\,\textrm{nV}/\sqrt{\textrm{Hz}} \times (1\,\textrm{Hz/f})^{1/2}$ is required to not degrade the performance measured with short circuited actuator. To reach the calculated thermal noise limited performance, the voltage noise even has to be below $23\,\textrm{nV}/\sqrt{\textrm{Hz}} \times (1\,\textrm{Hz/f})^{1/2}$. These are challenging specifications, especially in the low frequency range.

\begin{figure}
		\centering\resizebox{0.48\textwidth}{!}{%
		\includegraphics{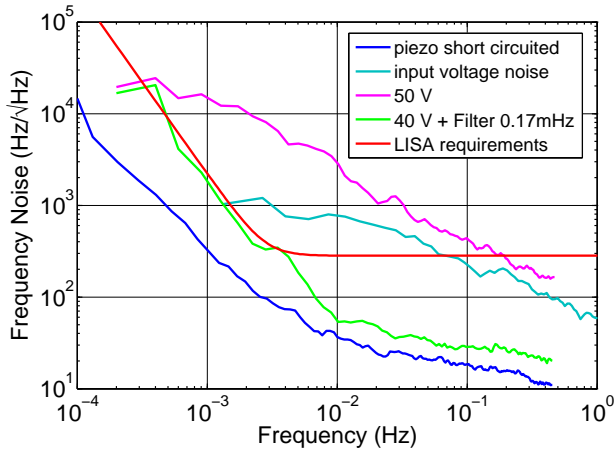}}
		\caption{ASD of the PZT stack piezo cavity driven with the PA 45 amplifier. The amplifier was operated with an amplification of 10 in each case. For measuring the input voltage noise the input of the amplifier was short circuited.}
		\label{fig:voltage_noise}
	\end{figure}

For building piezo control electronics, a low noise voltage reference and low noise amplifiers are required. The performance of different voltage references in the low frequency range has already been investigated \cite{FTS09}. The best such references can reach a relative noise level of less than $10^{-6}\,/\sqrt{\textrm{Hz}}$ at $0.1\,$mHz. We used a REF02D voltage reference (Analog Devices), which turned out to be sufficient for our purpose.
In the low voltage range, we employed an inexpensive OP27G (Analog Devices) operational amplifier for which we measured no increase in frequency noise. For higher voltages, we used an PA45 (Apex) operational amplifier which can drive a piezo with up to $150\,\textrm{V}_\textrm{pp}$. This amplifier caused an increased frequency noise which, however, could be suppressed by a $0.16\,$mHz low pass filter (metallized polypropylene capacitor $10\,\mu$F, metal film resistors $10 \times 10\,\textrm{M} \Omega$) to below the LISA requirements (Fig. \ref{fig:voltage_noise}).

The application of a low pass filter prevents fast tuning of the cavity resonance frequency. To overcome this restriction, we added an AC coupled low voltage bypass path to the filter, which enables fast tuning of the cavity with a tuning range of several $10\,$MHz without increase in frequency noise. This frequency separating filter, thus, allows small but fast variations around a high offset voltage.

\subsection{Influence of a varying piezo driving voltage}

To test the effect of a slowly varying driving voltage, we applied a voltage ramp to the PZT stack cavity. While using the low pass filter described above, the voltage was slowly increased from $36\,$V to $43\,$V which resulted in a frequency change of $50\,$MHz. With a $0.2\,$mV/s voltage ramp (corresponding to $ 1.7\,$kHz/s) no degradation of the frequency stability was observed.

The same holds true when the piezo is driven with a sinusoidal signal. We measured the frequency stability while different sinusoidal voltages of amplitudes up to $4\,$mV and frequencies between $1\,$Hz and $120\,$kHz were applied to the actuator without filtering. When the modulation frequency is higher than the lock bandwidth ($\sim30\,$kHz), the measured frequency noise is equal to the noise of the unmodulated cavity. When the modulation frequency is lower than the lock bandwidth, the noise spectrum has a peak at the modulation frequency, while the remaining spectrum shows no alteration.

Under the influence of a changing driving voltage, the piezo cavities show hysteresis and creep in a magnitude which is in accordance with manufacturer specifications. The position uncertainty due to hysteresis is up to 10\% of the nominal displacement and the logarithmically decreasing position change due to creep accounts for a few percent of a previous displacement. These non-linearities are an issue, when the piezo cavity is to be used in a feed forward configuration, i.e. when it is needed to set an exact frequency or to follow a specific frequency curve \cite{BBK10,JKL10}. In these cases, extensive modeling of the non-linear behavior would be necessary and it should be investigated how well the existing models (e.g. \cite{GoC97,JaK00}) can be adapted to a piezo cavity and how severe the remaining uncertainties are. However, these effects play a negligible role in applications where the piezo cavity is integrated in an outer feedback loop, as planned for LISA. Thus, such investigations are beyond the scope of this work. 


\section{Applications}

\subsection{Laser stabilization with modulated cavity}

A piezoelectrically tunable cavity can be used to implement a modified PDH locking scheme which employs an unmodulated laser and a modulated cavity. We tested this approach by stabilizing a laser to a piezo cavity (this measurement was done with the prestressed piezo cavity described later) which was modulated at $120\,$kHz. A laser stabilized this way shows no increase in frequency noise compared to a laser stabilized in the usual way. The advantage of this modified locking scheme is the availability of a tunable stable laser without modulation sidebands, which otherwise would only be possible using an external modulator.

\subsection{Use as transfer cavity}

Due to its tunability, a piezo cavity can be stabilized to another reference. In the LISA mission, for instance, it is planned to stabilize the piezo cavity to the 5 million kilometers long interferometer arms. To test the integration of the piezo cavity in such an outer feedback loop, we actively transferred the length stability of a highly stable fused silica (FS) cavity \cite{HSM09} to the PZT stack piezo cavity. This approach is quite similar to the widely used transfer cavity concept, where the stability of one laser is transferred to another using a piezo-tunable cavity \cite{BJM79,PAM06}. By locking the cavity to a reference laser and then stabilizing a target laser to this cavity, stabilities up to $10^{-11}$ have been transferred \cite{RAS10}. However, to our knowledge, the concept has not yet been realized with highly stable references in the $10^{-15}$ range.
	\begin{figure}
		\centering\resizebox{0.4\textwidth}{!}{%
		\includegraphics{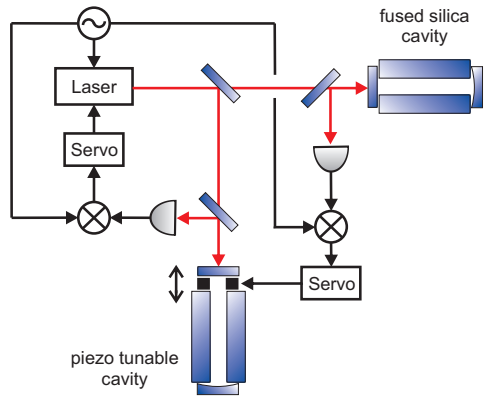}}
		\caption{Scheme of the stability transfer lock. The laser phase modulation is done with the piezo actuator which is attached to the laser crystal. Both Pound-Drever-Hall locks are realized with this modulation.}
		\label{fig:pzt_lock}
	\end{figure}

The stability transfer was realized with a laser that was coupled to both the piezo cavity and the fused silica cavity (Figure \ref{fig:pzt_lock}). First, the laser frequency was locked to the resonance of the piezo cavity using the PDH technique. Then, the piezo cavity was tuned to a resonance of the fused silica cavity, so that a PDH error signal can be observed by sweeping the voltage applied to the piezo cavity. This error signal was finally used for stabilization of the length and, thus, of the resonance frequency of the piezo cavity. For performing the piezo lock, the frequency separating filter described in the previous section was used. In this way, we succeeded in locking the resonance frequency of the piezo cavity to the fused silica reference cavity with a lock bandwidth of $5\,$kHz. The demands on the lock performance are here quite low, since the piezo cavity has such a good `free-running' stability.
	\begin{figure}
		\centering\resizebox{0.48\textwidth}{!}{%
		\includegraphics{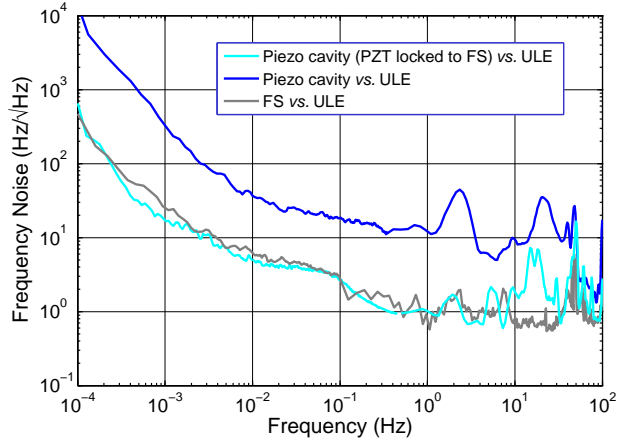}}
		\caption{Frequency noise of the externally stabilized piezo cavity. For comparison the frequency noise of the unstabilized piezo cavity and the fused silica (FS) reference are shown as well.}
		\label{fig:asd_pzt_locked}
	\end{figure}

To determine the frequency noise of the stabilized piezo cavity, a beat measurement with a laser stabilized to the ULE cavity was performed. Figure \ref{fig:asd_pzt_locked} shows the frequency noise of the stabilized and the non-stabilized piezo cavity as well as the noise of the fused silica cavity. It is apparent that the frequency noise of the piezo cavity is reduced by about one order of magnitude by stabilization, so that the stability of the fused silica cavity is nearly completely transferred to the piezo cavity. The remaining differences in seismic noise result from differing measurement conditions.

\section{Prestressed piezo cavity}

When a piezo cavity is to be used in mobile or space applications, it has to be considered that the brittle piezo ceramics cannot withstand high tensile or shear forces. Such forces do occur in dynamic operation of the actuator, but can also arise, e.g., during the launch of a space mission. They are usually handled by prestressing the piezo actuator. Since it is difficult to build highly stable cavities with commercially available prestressed actuators, we designed a piezo-tunable cavity where the piezo actuator is clamped between the cavity spacer parts themselves and is, thus, prestressed.

\subsection{Concept and assembly}

\begin{figure}
		\centering\resizebox{0.45\textwidth}{!}{%
		\includegraphics{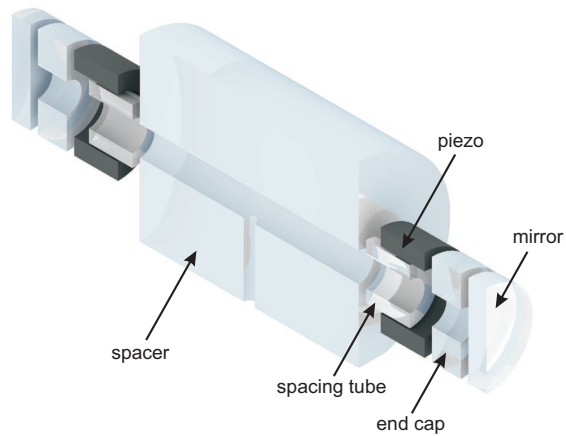}}
		\caption{Model of the prestressed piezo cavity (exploded view). The PZT stack actuators are $25\,$mm in diameter and have a length of $12\,$mm. The spacing tubes are $0.5\,\mu$m shorter than the actuators and have a diameter of $15\,$mm and a wall thickness of $3.5\,$mm.}
		\label{fig:prestressed}
	\end{figure}

Figure \ref{fig:prestressed} shows a model of the prestressed piezo cavity.  
It consists of a spacer, two spacing tubes, and two end caps made from Zerodur, as well as fused silica mirrors and PZT stack actuators. The special feature of the design is that the spacing tube located inside the piezo actuator is $0.5\,\mu$m shorter than the piezo itself, which leads to the aspired compression of the piezo actuator.

Both spacing tubes were bonded to the spacer and to the corresponding end cap, while the piezo actuators were contracted by continuously applying a negative voltage. Thus, after removal of this negative voltage, the expanded actuators are prestressed while the spacing tube is stretched. Tuning of the prestressed piezo cavity with positive voltage corresponds to further stretching the spacing tube.

The bonding was realized with hydroxide-catalysis bonding \cite{Gwo2001,Gwo2003} where silicate like networks are created between the surfaces. This bonding technique has the advantage of a high bonding strength and, especially important for this assembly, a small bond thickness $<100\,$nm.

\subsection{Calculations}
The dimensions of the spacing tube are chosen such that the resulting spring force equals roughly the spring force of the piezo actuator. This leads to a reduction of the maximum displacement of the actuator by a factor of $\sim 2$. With the machined length difference of $0.5\,\mu$m the piezo is subjected to a prestress of $0.7\,$MPa after bonding. The spacing tube is meanwhile stretched, which results in an internal stress of $1.7\,$MPa. This is well below the tensile strength of Zerodur, which is shown in recent measurement to be above $40\,$MPa \cite{HND09}.

For tuning the cavity over one free spectral range of $1.5\,$GHz a voltage of $\sim 100\,$V is required. At this displacement, the spacing tubes are subjected to a tensile stress of $3.7\,$MPa and the endcaps and mirrors are deformed due to the expanded piezo. This effect was simulated with a finite element method (FEM) and the end cap thickness was chosen such that the deflection is smaller than $20\,$nm and the deformation of the cavity mode geometry can be neglected.

\subsection{Cavity properties and measurements}

The $100\,$mm long prestressed cavity has a free spectral range of $1.5$\,GHz. The cavity was installed in the same setup as the other tested cavities. We obtained a coupling efficency of $70\%$ and measured a cavity resonance linewidth of $8\,$kHz. The tuning coefficient is $16\,$MHz/V. We tuned the resonance frequency of the cavity over more than one free spectral range by applying a voltage of $100\,$V. It was, thus, demonstrated that, in accordance with the calculations, material and bond can bear this stress without failure.

\begin{figure*}
		\centering\resizebox{0.98\textwidth}{!}{%
		\includegraphics{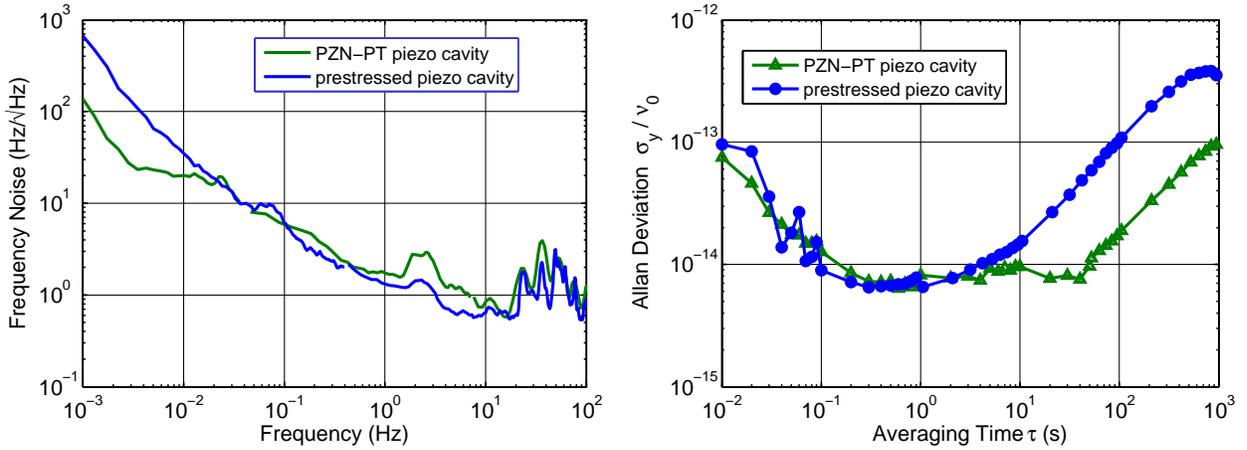}}
		\caption{ASD and relative Allan deviation of the prestressed piezo cavity in comparison to the cavity with PZN-PT actuator.}
		\label{fig:noise_prestressed}
	\end{figure*}

The noise performance of the short circuited prestressed piezo cavity is shown in Fig. \ref{fig:noise_prestressed}. The cavity has almost the same white and flicker noise as the other tested piezo cavities. There are only differences in the random walk noise of the prestressed piezo cavity which is increased by a factor of about 5. This is presumably caused by relaxations of the cavity materials, which have a relatively high internal stress due to the prestressed configuration.  

\subsection{Possible application in intra cavity spectroscopy}

Our prestressed piezo cavity design could be used for frequency references based on intra cavity spectroscopy of molecular or atomic gases where the cavity has to be tuned (and locked) to the optical transition of interest \cite{MYD99,YMH00,CMD04}. In particular, our cavity features a clean, hermetically sealable, chemically inert intra cavity volume. This is of advantage to maintain the purity of the spectroscopic gas, especially for reactive species. At the same time, the piezo actuator is protected from corrosive gases and the danger of electrical flash-overs at unfavorable pressures is avoided.

\section{Conclusion}

We demonstrated that with different piezoelectrically tunable cavities a frequency noise $<30\,\textrm{Hz}/\sqrt{\textrm{Hz}}$ for frequencies $>4\,$mHz and a relative frequency stability better than $1\cdot 10^{-14}$ at integration times from $1\,$s to $20\,$s can be achieved.  We observed no significant differences in frequency stability of the cavities with PZT stack, PZT tube or PZN-PT actuator within our measurement precision and the performance is most likely not limited by the piezo actuators. The cavities built with these actuators can be tuned over more than one free spectral range, so that the laser can be stabilized to any frequency of interest. For such a broad tuning range, a high voltage power supply is required, which needs to be filtered to not degrade the noise performance. Fast tuning of the cavity without increase in frequency noise is possible with a tuning range of several $10\,$MHz. Furthermore, a piezo cavity where the piezo is prestressed by the cavity spacer components to handle high forces has been realized. It shows a similar noise performance and tuning properties as the piezo cavities without prestressing. Piezo-tunable cavities are suitable for the use in the gravitational wave observatory LISA and are interesting devices for other applications where an optical frequency reference which is stable and tunable at the same time is desired.

\section*{Acknowledgement}

We thank the AEI Hannover and especially Johanna Bogenstahl for providing the expertise and facilities for the hydroxide-bonding procedure. This work is supported by the German Space Agency DLR with funds provided by the Federal Ministry of Economics and Technology (BMWi) under grant number DLR 50 OQ 0601. E.V. Kovalchuk is also associated with the Frequency Standards Laboratory, P. N. Lebedev Physical Institute, Moscow, Russia.


%
%

%

\end{document}